# Slow protein fluctuations explain the emergence of growth phenotypes and persistence in clonal bacterial populations

Andrea Rocco, Andrzej M. Kierzek, Johnjoe McFadden

*Faculty of Health and Medical Sciences, University of Surrey, Guildford, United Kingdom*

## Abstract

One of the most challenging problems in microbiology is to understand how a small fraction of microbes that resists killing by antibiotics can emerge in a population of genetically identical cells, the phenomenon known as persistence or drug tolerance. Its characteristic signature is the biphasic kill curve, whereby microbes exposed to a bactericidal agent are initially killed very rapidly but then much more slowly. Here we relate this problem to the more general problem of understanding the emergence of distinct growth phenotypes in clonal populations. We address the problem mathematically by adopting the framework of the phenomenon of so-called weak ergodicity breaking, well known in dynamical physical systems, which we extend to the biological context. We show analytically and by direct stochastic simulations that distinct growth phenotypes can emerge as a consequence of slow-down of stochastic fluctuations in the expression of a gene controlling growth rate. In the regime of fast gene transcription, the system is ergodic, the growth rate distribution is unimodal, and accounts for one phenotype only. In contrast, at slow transcription and fast translation, weakly non-ergodic components emerge, the population distribution of growth rates becomes bimodal, and two distinct growth phenotypes are identified. When coupled to the well-established growth rate dependence of antibiotic killing, this model describes the observed fast and slow killing phases, and reproduces much of the phenomenology of bacterial persistence. The model has major implications for efforts to develop control strategies for persistent infections.

## Introduction

The phenotypic heterogeneity observed in populations of genetically identical cells is a ubiquitous and intriguing phenomenon, whose precise origin is still far from being fully understood. Since it appears to play a relevant role in many different contexts, ranging from the emergence of drug tolerance phenotypes in bacterial populations [1], to the somatic evolution of cancer cells [2], to cell differentiation [3], it is believed that it does not emerge by mere accident, but it is rather the result of maybe complex gene regulatory processes.

Stochastic processes are at the base of phenotypic heterogeneity [4]. It is conceivable that different noise sources, both static and dynamic, play a different role in the emergence of heterogeneous phenotypes in clonal populations of cells. Extrinsic noise sources, such as for instance fluctuations in the number of ribosomes or RNA polymerases, are static, and therefore are often the best candidates when looking for mechanisms that produce different cell phenotypes. In contrast, so-called intrinsic noise, related for instance to the bursting activity of gene expression or to the repartition of protein molecules in daughter cells at cell division, creates typically fast dynamical fluctuations, and therefore no stable phenotypes can emerge [5].

In this paper we focus on intrinsic noise, and propose a specific mechanism that slows down the intrinsic fluctuations associated with gene expression and protein repartition during cell division. We show that this mechanism is in fact sufficient to account for the emergence of phenotypic heterogeneity in clonal populations.

In order to do this, we make reference to the concepts of ergodicity breaking and epigenetic landscape. Ergodicity breaking [6] is a concept borrowed from dynamical systems theory and statistical physics, and recently suggested to play a role in biology as well [2]. It relies naturally on the notion of the epigenetic landscape, first proposed by Waddington in 1957 [7] in a developmental context.

Inspired by [7], we represent the cell state as a point in a multidimensional space (the so-called configuration space), whose axes correspond to the expression values of each gene of the cell. The specific gene network dynamics determines what gene configurations are accessible to the cell, and therefore restricts the cell to a limited set of possible states. Computing the inverse probability that the cell is found in any state [2], and plotting it on a further axis, defines a hyper-surface in the state space, which describes pictorially the network dynamics. This hyper-surface corresponds to the epigenetic landscape introduced in [7]. The cell explores the epigenetic landscape driven by the network dynamics, and by temporal stochastic fluctuations of genetic and non-genetic origin. In Fig. 1, we present a pictorial description of the probability distribution of the cell states and of the corresponding epigenetic landscape in the case of a single gene.

The epigenetic landscape plays the same role as the energy landscape in Hamiltonian system. Because of the probabilistic definition adopted here, the modes of the probability distribution of the different gene configurations correspond to landscape minima, and probability minima correspond in turn to local maxima of the landscape. In the same way as a Hamiltonian system relaxes in time towards the energy minimum, in the present picture the cell tends to approach the minima of the epigenetic landscape, which can then be called metastable, or attractor, states. However, since conceptually derived from the knowledge of the probability distribution of cell states, the epigenetic landscape combines both deterministic and stochastic components of the dynamics, and metastable states do not necessarily correspond to stationary states of the underlying deterministic dynamics only.

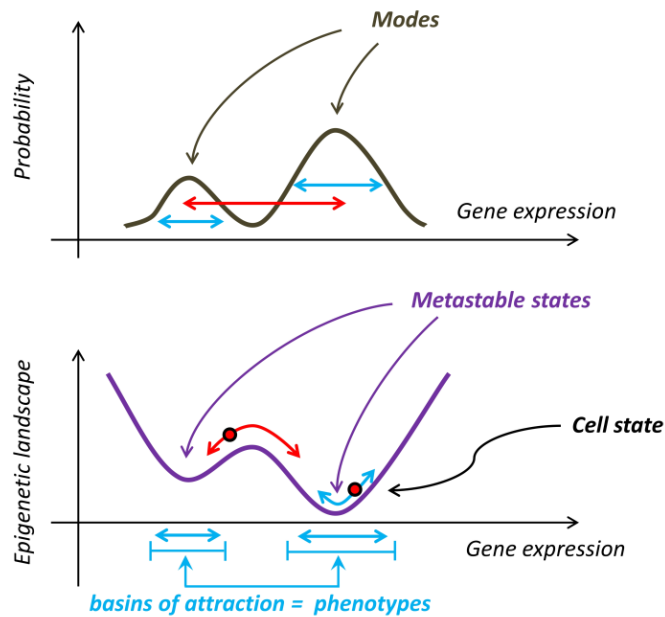

**Fig. 1. The epigenetic landscape.** The probability distribution (upper panel) and the corresponding epigenetic landscape (lower panel) are shown for a 1-dimensional system, characterized by the expression levels of one single gene. The bimodal structure of the probability distribution of the cell states induces a dual landscape, similar to the potential energy landscape in Hamiltonian systems, in which the modes of the probability distribution are mapped into metastable states. The profile of the landscape is a manifestation of the gene dynamics, which the cell can explore at equilibrium, driven by stochastic fluctuations. For fluctuations small enough (**light blue arrows**), the cell remains confined in the basin of attraction around each metastable state, while strong enough fluctuations (**red arrows**) will make the cell hop from one basin of attraction to an adjacent one. If these transitions are rare, and the sojourn time within a basin of attraction is comparable with the observational time, we identify all possible states within that basin of attraction as one single (noisy) phenotype.

Complex landscapes, possibly characterized by many maxima and minima, and hills and valleys, are likely in gene regulatory networks because of the ubiquitous existence of gene feedback circuits [8]. The set of cell states belonging to the same valley, and relaxing toward the corresponding metastable state is commonly called the basin of attraction of that metastable state.

The notion of basin of attraction suggests a useful definition of a phenotype. We propose to interpret the basin of attraction of each metastable state as one phenotype. Namely, we associate all cell states within the basin of attraction of a given metastable state to the same phenotype, and states belonging to basins of attraction of different metastable states to distinct phenotypes (Fig. 1). The case of only one metastable state in the system corresponds trivially to one single phenotype. In the following we are interested instead in the case when multiple metastable states are present.

At equilibrium, stochastic fluctuations are responsible for the wandering of the cell state in the landscape, both within basins of attractions when fluctuations are small, and across them, when fluctuations are large. In the first case fluctuations do not modify the cell phenotype, while in the second case a phenotypic change is produced. We define the permanence or sojourn time $\tau_p$ as the average time the cell has to wait before being exposed to a fluctuation large enough to make a "hop" from the basin of attraction of one metastable state to an adjacent one. Furthermore, we call observational time $\tau_{\exp}$ the time over which a typical experiment or observation is performed. It is then natural to refine our definition of a phenotype, and interpret the basins of attraction of distinct metastable states as distinct 'observable' phenotypes only if the permanence time is larger or at least of the order of the observational time, namely if $\tau_p / \tau_{\exp} \geq 1$.

In the case when the permanence time is considerably shorter than the observational time, $\tau_p / \tau_{\exp} \ll 1$, the system hops rapidly among the basins of attraction of the available metastable states, and the observed time-averaged behaviour is the same for any observed cell. For this reason, the time average of the relevant variable (for instance a specific protein concentration) for a single cell equals the ensemble average over the population of cells. In this case the system is said to be in the ergodic phase [6]. The fact that all cells behave the same during the observation leads to conclude that only one (average) phenotype is present in the population, despite the presence of multiple metastable states, with multiple basins of attraction.

In contrast, when the permanence time is large, at least of the order of the observational time, each cell maintains its own individuality during the observation, and the time-averaged variable of interest measured over the observational time will differ from cell to cell. The time average differs now from the ensemble average, and in the limit of infinite observational time the system is said to be in the non-ergodic phase. This case allows for distinct phenotypes to become observable. The transition between ergodic and non-ergodic phases is called ergodicity breaking [6].

The standard definition of non-ergodic phase relies on the observational time going to infinity. This implies that cells belonging to the basin of attraction of one metastable state cannot access any other basin of attraction in any finite time. In the landscape picture introduced above, this would correspond to the barriers between different basins of attraction becoming of infinite height, and the basins of attraction becoming dynamically disconnected. The whole state space would then be partitioned into distinct islands, no matter how large the fluctuations would be. The impossibility for the system to explore the whole space in any finite time would then imply impossibility of reaching equilibrium.

In fact this definition of non-ergodic behaviour is too strong for our purposes, since much of the interesting biological dynamics may happen in an off-equilibrium regime, well before the system has had the time to attempt to explore the whole phase space. Furthermore, strict ergodicity breaking would not be verifiable experimentally, because observational times are anyway experimentally finite. For this reason, we use the concept of ergodicity breaking in a weaker way, to indicate that the system appears as non-ergodic when measured over a finite observational time, which is generically large but finite, dictated by the experiments, and define the weakly non-ergodic phase by the same condition used in our definition of a phenotype, namely $\tau_p/\tau_{\exp} \geq 1$. In doing so, we leave open the possibility that cells might eventually be able to visit all available states, and therefore equilibrate, on infinite observational times. Our definition of weak ergodicity breaking is similar in spirit to the one introduced in [9,10] for disordered systems, with the difference that we adopt a finite observational time.

The search for distinct phenotypes becomes then equivalent to searching for mechanisms capable of slowing down the dynamics, so as to increase the permanence time, and approximate the behaviour of the system as weakly non-ergodic, in the sense specified above. Slow-down of dynamics may be the result of different causes. For instance both the topological properties of the landscape, (namely the depth of the basins of attractions, or their internal possibly rugged structure), and the intensity and rapidity of the temporal stochastic fluctuations in gene expression are all factors expected to play a role. If temporal fluctuations in protein numbers are fast, in particular because of the relatively short cell division time that provides an efficient mixing mechanism of protein levels across generations, these contribute ergodic components to the full dynamics. In [11] for instance, in a developmental context, it is in fact hypothesized that the complexity of the landscape plays a major role. The Authors of [11] show that a complex rugged landscape emerges because of the complex multidimensional network of gene interactions. This implies the existence of high dimensional attractor states, and thus leads to the appearance of a relatively limited number of long-lived macroscopic states, which are interpreted as distinct phenotypes.

To explore how ergodicity breaking may generate distinct phenotypes we consider how it may emerge from protein control of cellular growth rate. Many factors, both genetic and environmental, do of course influence cellular growth rate [12,13]. However we here envisage the situation in which growth is inhibited by a protein and consider how distinct growth phenotypes may emerge due to a slow-down of protein fluctuations. In our model accumulation of protein in the cell leads to increasingly longer cellular division times, and therefore decreases the effectiveness of cellular division as a randomization process responsible for protein levels mixing [14]. Other processes responsible for mixing (transcription and protein degradation) are also kept at minimal efficiency, by assuming low gene expression and degradation.

In the Results section of the paper, we show that the system appears as non-ergodic, in the weak sense defined above, and is characterized by growth heterogeneities, which turn out to be stable over typical observational times. We analyse the model in the ergodic and weakly non-ergodic regimes, and show that bimodal distributions of growth rate are expected in the non-ergodic regime, for fast enough translation.

We also apply this model to considering how the resulting bimodal distribution of cellular growth rate may impact on downstream drug tolerance phenotypes. Growth rate is an important determinant of the response of cells to numerous stimuli including stresses such as starvation and exposure to toxins, drugs and biocidal agents [15–17]. We illustrate this effect by extending the model to examine killing of bacteria by antibiotics. We demonstrate that biphasic killing, a key characteristic of the enigmatic phenomenon of bacterial persistence, emerges in the weakly non-ergodic regime. We also present direct stochastic simulations (Gillespie), which support the analysis for the emergence of both growth phenotypes and persistence. In the Methods section we give details on our extension of the Gillespie algorithm to include protein controlled cellular division times.

## Results

### Bacterial Persistence

The term 'persisters' was first used by Bigger [18] to describe the ability of a small fraction of a population of genetically identical (isogenic) cells of *Staphylococcus aureus* to survive prolonged exposure to bactericidal concentrations of penicillin. Since then, the phenomenon has been described in nearly all known microbes and considered to be largely responsible for the resistance to antibiotic therapy of many chronic bacterial infections, such as tuberculosis (TB) [1] and in the resistance of biofilms to microbiocides and antibiotics [19,20]. The key signature of persistence is the biphasic kill curve obtained when bacteria in batch culture are exposed to a bactericidal antibiotic [1]: the killing rate is initially very high but then slows and may even level off to zero. Numerous factors have been proposed to be responsible for persistence but a landmark study in 2004 [21] examined antibiotic killing of *hipA7 E. coli* at the single cell level and demonstrated that persister cells were either slow-growing or non-growing at the time of antibiotic administration. The authors introduced a persistence model based on the simultaneous existence of two preexisting subpopulations consisting of normal and persistent cells, and a constant rate of stochastic phenotypic switching between the two cell types. The *hipA* gene was subsequently shown to encode the toxin component of a toxin-antitoxin (TA) module, *hipAB* [22] whose over-expression was shown to slow growth. It has recently been proposed that stochastic [23] or growth-rate mediated gene expression feedback mechanisms [24] in the regulatory circuits controlling expression of HipA cause bistability and switching between drug-sensitive normal and drug-tolerant persister states. However, drug-tolerance and persistence are arbitrarily assigned to the normal and persister cells in these models, rather than derived from the models.

Toxin components of toxin-antitoxin systems, such as HipA, are generally expressed at low level and, at sublethal concentrations, inhibit cell growth [25]. This suggests the development of a model based on the weak expression of a growth controlling protein, whose distribution may be subject to ergodicity breaking. However, the mechanism proposed here is not specific to toxin components, but may hold for any growth inhibiting protein. We then extend this model to include antibiotic-mediated killing of an isogenic bacterial population by assuming that the killing rate is proportional to the growth rate

[15], and examine whether ergodicity breaking may be involved in the phenomenon of persistence.

### The growth model

Let us consider then the behaviour of a system in which expression of a single gene controls cellular division times. Specifically, we consider the action of a gene that inhibits growth, and express the protein dependent cellular growth rate $g(p(t))$ in terms of a Hill function with unity Hill coefficient:

$$g(p(t)) = \frac{g_0}{1 + \kappa p(t)}. \tag{1}$$

Here $p(t)$ is the time dependent protein concentration, $g_0 = \ln 2/T_0$ is the maximal cellular growth rate, with $T_0$ being the zero protein division time, and $\kappa$ is a parameter that quantifies the growth-inhibitory strength of the inhibitory protein. The protein concentration is defined as $p(t) = n(t)/V(t)$, where $n(t)$ and $V(t)$ are respectively the number of protein molecules and the cellular volume at time $t$.

We then assume that the cellular volume $V$ satisfies the equation:

$$\frac{dV(t)}{dt} = \frac{g_0}{1 + \kappa p(t)} V(t). \tag{2}$$

For $\kappa = 0$, Eq. (1) reduces to the protein independent growth rate $g_0$, and Eq. (2) reproduces an exponential growth law for the cellular volume. Although it has been claimed that growth of cellular volume is linear in some systems [26], this has been questioned by later studies [27], and most recent modelling [28,29] has assumed exponential growth, which is also assumed here.

To take into account production of the inhibitory protein, we assume the following model of gene expression [28,29],

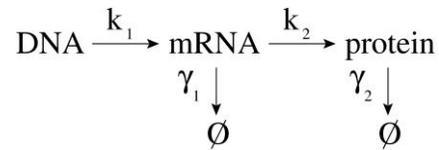

where the parameters $k_1$ and $k_2$ are respectively transcription and translation rates, while $\gamma_1$ and $\gamma_2$ describe degradation of mRNA and protein. Furthermore, cell division is implemented by imposing that cells divide when the cellular volume doubles. At division we make the assumption that the protein content of each mother cell is distributed binomially into the two daughter cells. Because of this process, and because of the bursting activity of gene expression, this model is intrinsically stochastic.

If the function $p(t)$ is known (for instance through direct stochastic simulations), we can formally solve Eq. (2) as

$$V(t) = V_0 \exp\left(\int_0^t g(p(t'))dt'\right). \quad (3)$$

The condition $V(T_{div}) = 2V_0$ leads then to the following implicit definition of the division time $T_{div}$:

$$1 = \int_0^{T_{div}} \frac{1}{T_0(1+\kappa p(t'))}dt'. \quad (4)$$

This expression can be solved explicitly for $T_{div}$ only if the stochastic variable $p(t)$ is known. If this is not the case, the integral in (4) cannot be computed, and we have to rely on approximation methods. In particular, we can find an approximate solution of (4) when the fluctuations of $p(t)$ are either very fast or very slow with respect to the cell cycle.

Fast gene expression fluctuations – The ergodic regime
If $p(t)$ fluctuates fast over $T_{div}$, we can replace $p(t)$ in (4) with its time average,

$$\bar{p} = \frac{1}{T_{div}} \int_0^{T_{div}} p(t')dt'. \quad (5)$$

Bursting activity and protein degradation are the stochastic processes responsible for protein fluctuations within generations, with protein reshuffling at cell division contributing further stochasticity across generations. In terms of the parameters of the model, fast gene expression fluctuations can be realized by assuming fast bursting activity (namely $k_1$ large) and fast protein degradation (namely $\gamma_2$ large, even though always smaller than the mRNA degradation rate $\gamma_1$ [28,29]). Protein reshuffling due to cell division is not expected to play a role in this regime, because subsequent randomization due to bursting and degradation will quickly decorrelate the protein content from its initial value, set just after cell division. As a consequence the value $\bar{p}$ as given by (5) will be the same across different generations, namely conserved within, and also across, cell lines.

Using then the ergodic hypothesis, $\bar{p} = \langle p \rangle$, with $\langle p \rangle$ the average over the cell population, leads to

$$T_{div} \approx T_0(1+\kappa\langle p \rangle). \quad (6)$$

By following [28,29], we then write the master equation associated with the processes above as

$$\frac{\partial w(p,t)}{\partial t} = \frac{\partial}{\partial p}\left[\left(\gamma_2 + \frac{\ln 2}{T_0(1+\kappa\langle p\rangle)}\right)pw(p)\right] + k_1 \int_0^p K(p-p')w(p')dp', \quad (7)$$

where $w(p)$ is the protein distribution over the population, the first two terms in the right hand side represent dilution effects due to protein degradation and cell division, and the last term is protein production, with $K(p) = (1/b)\exp(-p/b) - \delta(p)$. Here $b$ is the average burst size during translation, and the Dirac delta function represents transitions away of $p$ [28,29].

An analytic stationary solution of the master equation (7) can be computed. This results in the Gamma distribution

$$w(p) = \frac{1}{b^a \Gamma(a)} p^{a-1} e^{-p/b}, \quad (8)$$

where $a = k_1/(\gamma_2 + \ln 2/(T_0(1+\kappa\langle p\rangle)))$ is the mean number of transcriptional bursts per cell cycle, $b = k_2/\gamma_1$ is the mean number of protein molecules produced per burst during translation, and $\Gamma(a)$ is the gamma function. In particular, by using $\langle p\rangle = ab$ we obtain the following expression for $a$:

$$a = \frac{\kappa k_1 k_2 - \gamma_1\gamma_2 - \gamma_1 \ln 2/T_0 + \sqrt{(\kappa k_1 k_2 - \gamma_1\gamma_2 - \gamma_1 \ln 2/T)^2 + 4\kappa\gamma_1\gamma_2 k_1 k_2}}{2\kappa\gamma_2 k_2}. \quad (9)$$

Since $T_{div}/a$ is the mean time between successive bursts, the condition of fast protein fluctuations over the cell cycle $T_{div}$, used to solve (4), requires $T_{div}/a < T_{div}$, namely $a > 1$. Since the observation time $\tau_{exp}$ is supposed to be larger than $T_{div}$, $a > 1$ guarantees ergodicity.

To assess the validity of our analytical predictions we performed direct stochastic simulations. We extended the Gillespie algorithm [30] to incorporate a cell division model where the division time is dependent on a growth controlling protein. We give details of the adapted simulation algorithm in the Methods section.

The resulting simulation data are reproduced very well by the Gamma distribution (8), with no fitting parameters (Fig. 2). The value for the parameter $a$ is computed according to Eq. (9), while we set $b = k_2 \ln 2/\gamma_1$, where the factor $\ln 2$ comes from averaging the volume over the cell cycle as in [28,29]. The strong agreement between analysis and simulations supports the validity of the ergodic hypothesis leading to eq. (6), and therefore to (9). In Fig. 2 we also show the corresponding division time and growth rates distributions, as resulting from the simulation. Notice that in this regime, each distribution is characterized by a single mode only, representing a single (noisy) phenotype. In the epigenetic landscape picture, this corresponds to a single valley.

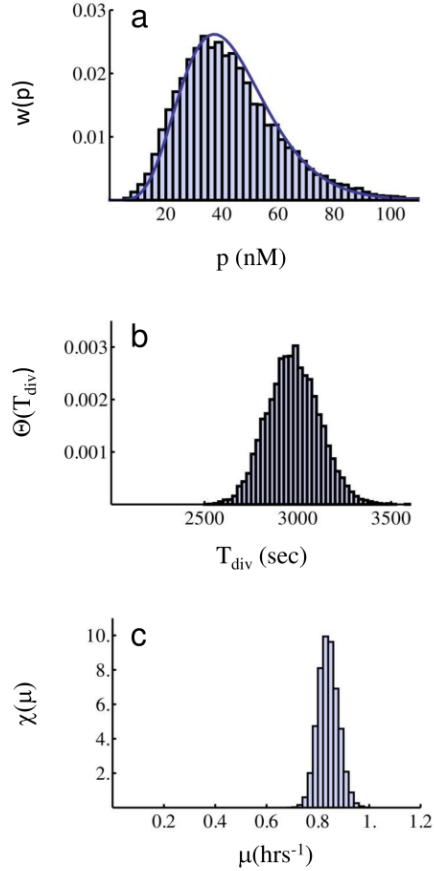

**Fig. 2. Protein, division time, and growth rate distributions in the ergodic regime.** Parameter values are $k_1=3\cdot10^{-2}$, $k_2=0.35$, $\gamma_1=0.04$, $\gamma_2=4\cdot10^{-3}$ (all units in sec$^{-1}$), $T_0=2100$ (sec), $\kappa=0.01$ (nM)$^{-1}$, $V_0=1.7$ fl. Histograms are the result of direct stochastic simulations (see text for details). The full curve in panel **(a)** corresponds to the Gamma distribution (8) with parameters $a=7.14$ as from (9) and $b=k_2\ln2/\gamma_1=6.06$, and no other fitting parameters. The ln2 here comes from averaging the cellular volume over the cell cycle [28,29]. Panel **(b)** shows the histogram for the division time distribution, obtained by direct measurement of $T_{div}$ during the simulation. The histogram for the growth rate distribution (panel **(c)**) is obtained from the measured $T_{div}$ by computing $\mu=\ln2/T_{div}$. In the ergodic regime, the growth rate distribution is characterized by one mode only, corresponding to one single phenotype.

## Slow gene expression fluctuations – The weakly non-ergodic regime

Let us consider now the case when $p(t)$ is slowly fluctuating during the cells' life span. In this case we consider $p(t)$ almost constant in (4), and obtain

$$T_{div} = T_{div}(p) \approx T_0(1 + \kappa p), \qquad (10)$$

with $p$ the protein concentration just after cell division. Slow fluctuations in gene expression over the cell cycle will now be produced by slow transcription ($k_1$ small) and slow protein degradation ($\gamma_2$ small). The requirement of slow transcription corresponds to imposing the condition $T_{div}/a > T_{div}$, which produces now $a < 1$, and guarantees that fluctuations in gene expression are slow over the cell cycle $T_{div}$. Furthermore the requirement of slow protein degradation over the cell cycle corresponds to imposing $1/\gamma_2 > T_{div}$. These two conditions together make Eq. (10) valid. However, in contrast to the fast fluctuations case, protein reshuffling at cell division will now play a role in randomizing protein levels, and resetting them across generations. This implies that the regime $k_1$ and $\gamma_2$ small does not guarantee in general that protein levels will be constant within cell lines. So, in general, slow fluctuations in gene expression and degradation are a necessary but not sufficient condition for (weak) ergodicity breaking. Weak ergodicity breaking will be realized by imposing the further requirement that the observational time be smaller than the division time, $\tau_{exp} < T_{div}$. This condition fixes the maximal length of an experiment aiming at detecting individually stable phenotypes.

If we now supplement the regime of $k_1$ and $\gamma_2$ small with the further assumption of fast translation ($k_2$ large), any time a molecule of mRNA is produced, with high probability a large burst of protein molecules will be translated. Therefore in this regime all cells will undergo rare transcriptional events, from which however large amounts of protein are produced. As a consequence cells will be most likely to fall in one of the two categories, either with close to zero protein content (because transcription is rare), or with a large amount of protein (because translation is very efficient). The number of cells showing an intermediate amount of protein numbers is then relatively negligible in this regime.

It should be noted that this regime, with fast translation, reproduces the features of the weak ergodicity breaking defined above. The two portions of phase space respectively characterized by negligible and very large protein contents appear to be weakly connected phase space islands, with negligible transition probabilities between them over large but finite observational times.

Within the epigenetic landscape picture presented in the Introduction, the translation rate $k_2$ can then be regarded as a parameter that controls the landscape morphology, by inducing a transition from a single well to a double-well in the growth rates landscape. In the dual representation in terms of probability distribution, this situation will correspond to the emergence of a bimodal probability distribution for growth rates, which is then the result of a weak ergodicity breaking. The role of the parameter $k_2$ appears to be that of "separating" the non-ergodic components of the system.

In the weakly non-ergodic regime, we expect the Gamma distribution (8) to be still a (approximate) solution of the model. In fact the slow fluctuations at the protein level correspond formally to slow varying heterogeneities in the corresponding mean number of bursts $a$ and mean burst size $b$. No matter what the distributions of these heterogeneities are, these can be integrated over, and produce again an approximate Gamma distribution, as shown in [31]. For this reason, we make the well justified assumption that in the weakly non-ergodic regime the protein is still distributed with a Gamma distribution, for which we will evaluate the corresponding $a$ and $b$ values numerically from our stochastic simulations.

As a result, in the weakly non-ergodic regime the population structure can be represented in terms of a continuum of subpopulations, which are virtually non-interacting because of the limited mixing among different protein levels. The growth dynamics of each subpopulation is thus defined as

$$\frac{dX_p}{dt} = \frac{\mu_0}{1+\kappa p} X_p(t), \qquad p \in [0,\infty] \qquad (11)$$

Here $X_p(t)$ represents the number of cells in the subpopulation characterized by protein content $p$, and $\mu_0$ is the maximal population growth rate, identical in value to $g_0$. The protein $p$ is distributed as a Gamma distribution.

The non-interacting population dynamics (11) is an approximation to the real dynamics based on neglecting mixing terms among different protein levels. This approximation is valid for times smaller than any mixing time scale in the system, for which the division time is a lower bound, as discussed. For longer times, mixing will become effective, the system will restore ergodicity and equilibrate, and single cells will lose their phenotypic individuality.

By using (10), and the fact that $|\Theta(T_{div})dT_{div}| = |w(p)dp|$, $T_{div}$ is distributed as

$$\Theta(T_{div}) = \frac{1}{(\kappa b)^a \Gamma(a)} \frac{1}{T_0} \left(\frac{T_{div}}{T_0} - 1\right)^{a-1} e^{-\frac{1}{\kappa b}\left(\frac{T_{div}}{T_0} - 1\right)}. \qquad (12)$$

and by using the non-interacting population approximation leading to Eq. (11), the distribution of the $p$ dependent cellular growth rates $\mu(p) = \mu_0/(1+\kappa p)$ defined by (11) results in:

$$\chi(\mu) = \frac{\mu_0}{(\kappa b)^a \Gamma(a)} \frac{1}{\mu^2} \left(\frac{\mu_0}{\mu} - 1\right)^{a-1} e^{-\frac{1}{\kappa b}\left(\frac{\mu_0}{\mu} - 1\right)}. \qquad (13)$$

It can be proven that for $a < 1$ and $(1+a+1/\kappa b)^2 - 8/\kappa b > 0$ (see Methods section), the distribution $\chi(\mu)$ is characterized by two modes: a first mode at slow growth rates associated with cells expressing high values of protein, and a second mode at the maximal growth rate when the majority of cells present negligible protein concentration. In the weakly non-ergodic phase this model generates a bimodal dynamics in a system which does not assume *apriori* the two phenotypic states associated with two pre-existing subpopulations (slowly and fast growing cells). The stochastic effects stemming from the individual cells' gene expression, together with the chosen protein control of division times, are in fact solely responsible for the appearance of the growth heterogeneity in the population.

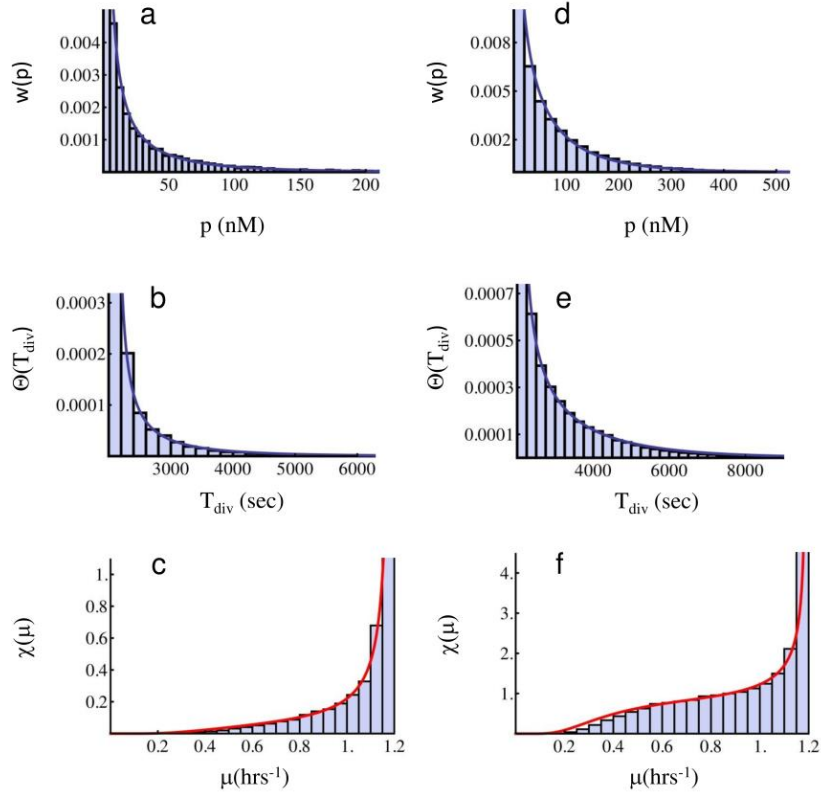

**Fig. 3. Protein, division time, and growth rate distributions in the slow fluctuations case**. Parameter values for the simulation (**a,b,c**) are $k_1=1.6\cdot10^{-5}$, $k_2=1.0$, $\gamma_1=0.01$, $\gamma_2=4\cdot10^{-5}$ (all units in sec$^{-1}$), $T_0=2100$ (sec), $\kappa=0.01$ (nM)$^{-1}$, $V_0=1.7$ fl. The full curve in panel (**a**) corresponds to the Gamma distribution (8) with parameters $a=0.045$ and $b=96.78$ fitted as described in the text. The full curves in panels (**b**) and (**c**) correspond respectively to the division time distribution, Eq. (12), and growth rate distribution, Eq. (13), evaluated with the same parameters. Parameter values for the simulation (**d,e,f**) are $k_1=1.6\cdot10^{-4}$, $k_2=1.0$, $\gamma_1=0.01$, $\gamma_2=4\cdot10^{-5}$ (all units in sec$^{-1}$), $T_0=2100$ (sec), $\kappa=0.01$ (nM)$^{-1}$, $V_0=1.7$ fl. The full curve in panel (**d**) corresponds again to the Gamma distribution (8) with parameters $a=0.5$ and $b=112.68$ fitted as described in the text, and the full curves in panels (**e**) and (**f**) correspond to Eqs. (12) and (13), evaluated with these same parameters. The good agreement between direct simulations and the predictions (12) and (13) supports the validity of the slow fluctuations approximation, leading to Eq. (10). The peak on the right in the growth rate distribution corresponds to the majority of cells growing at the maximal growth rate.

In Fig. 3 we show Gillespie simulations of protein, division time, and growth rate distributions for two different parameter sets, both determining slow fluctuations. In this case, theoretical predictions of the mean number of bursts $a$ and mean burst size $b$ are not available. Therefore we estimated their values by measuring the first and second moment of the simulated data, and by using $a = \langle p \rangle^2 / \sigma^2$ and $b = \sigma^2 / \langle p \rangle$, consistent with the assumption of an underlying Gamma distribution, and with $\sigma$ being the variance of the data. In panels (a) and (d) of Fig. 3, the corresponding protein Gamma distributions $w(p)$ are shown, which fit very well the simulations. In panels (b) and (e) we instead compare the distribution of division times $T_{div}$ directly measured from the simulation with the theoretical prediction $\Theta(T_{div})$ given by Eq. (12), using the same $a$ and $b$ values estimated from the protein distribution. These same parameter values are also used to compare the growth rate data, obtained from the measured $T_{div}$'s by computing $\mu = \ln 2 / T_{div}$, with the theoretical growth rate distribution $\chi(\mu)$, Eq. (13). We show this comparison in the panels (c) and (f) of Fig. 3. Even though mixing between different subpopulations due to cell division is expected to play a role, the agreement between simulations and the theoretical predictions for the division time distribution (12) and the growth rate distribution (13) is excellent. The comparison shown in panels (b) and (e) supports the validity of (10), while panels (c) and (f) support also the non-interacting population dynamics (11). This in turn shows that non-ergodic components dominate the full dynamics.

At fast translation, we make the same comparison, again for the three distributions, and show the result in Fig. 4. Again, the parameters $a$ and $b$ are estimated from the protein data, and their values are used in the protein Gamma distribution $w(p)$, Eq. (8), in the division time distribution $\Theta(T_{div})$, Eq. (12), and in the growth rate distribution $\chi(\mu)$, Eq. (13). Also in this case the agreement between analysis and simulations is excellent, and supports the validity of Eq. (10), based on slow fluctuations, and the weakly non-ergodic regime for large $b$ values. The peak to the left in panel (c) corresponds to a small subpopulation representing slowly growing cells, and includes all cells in the tail of the protein Gamma distribution illustrated in panel (a). The peak on the right corresponds instead to the majority of cells in the population, characterized by zero or negligible protein content, and therefore growing at the maximal growth rate.

It is interesting to evaluate the different mixing time scales for the set of parameters used for the simulation in Fig. 4. The bursting time scale results in $T_{div}/a \approx 1.5 \cdot 10^4$ (sec), the degradation time scale is $1/\gamma_2 = 2.5 \cdot 10^4$ (sec), and the typical division time can be estimated as $T_{div} = T_0(1 + \kappa ab) \approx 10^4$ (sec) (see caption of Fig. 4 for the corresponding parameter values). These values suggest that equilibration will take place for times much longer than the largest of these time scales, namely longer than $2.5 \cdot 10^4$ (sec), while phenotypic individuality will be maintained for times smaller than the smallest time scale, namely $10^4$ (sec). The sampling for constructing the distributions shown in Fig. 4 was then performed after $10^5$ sec of simulation, with further simulations with longer time runs (up to $10^{10}$ sec) before sampling not producing any appreciable change in the profile of the distributions (data not shown). In these conditions, the good agreement between the weakly non-ergodic assumption, represented by the non-interacting population dynamics, Eq. (11), and the simulation, shows that the dominant contribution to the population structure comes from non-ergodic components, with most cells conserving their own individuality, and performing only limited transitions between the two sub-populations with slow and fast growth rates. In this sense, these two subpopulations can be considered as non-interacting, and the resulting distribution is predominantly made of cells conserving their own growth rate. In the epigenetic landscape picture, this situation corresponds to a double-well landscape, with limited transitions between the two wells, and defines two distinct growth rate phenotypes. With the parameter as in Fig. 4, a conservative estimate for the duration of a typical single cell experiment aiming at observing distinct non-mixing phenotypes is of the order of $10^4$ sec. However given the validity of (11) well beyond this limit, most (not all) cells will remain in their own state for much longer times.

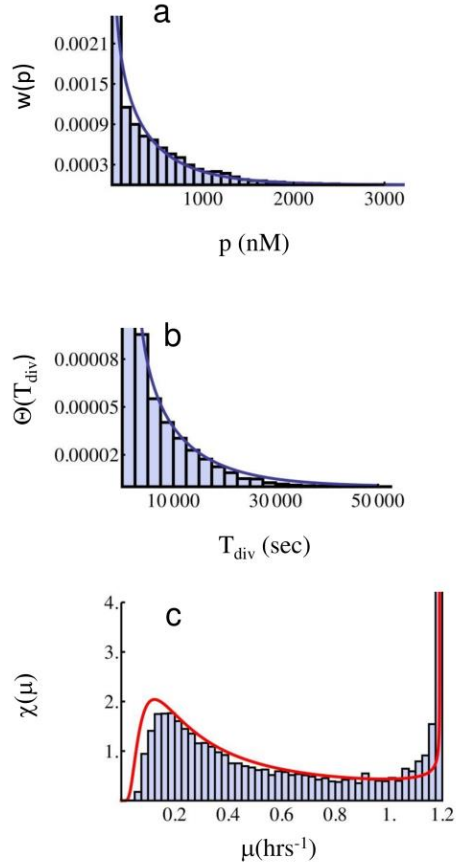

**Fig. 4. Protein, division time, and growth rate distributions in the slow fluctuations case and fast translation**. Parameter values for the simulation are $k_1=1.0 \cdot 10^{-4}$, $k_2=5.0$, $\gamma_1=0.01$, $\gamma_2=4 \cdot 10^{-5}$ (all units in sec$^{-1}$), $T_0=2100$ (sec), $\kappa =0.01$ (nM)$^{-1}$, $V_0=1.7$ fl. The full curve in panel **(a)** corresponds to the Gamma distribution (8) with parameters $a=0.69$ and $b=579.8$ fitted as described in the text. The full curves in panels **(b)** and **(c)** correspond to Eqs. (12) (division times distribution) and (13) (growth rate distribution) respectively, evaluated with the same parameters. In this parameter regime, weakly non-ergodic components dominate the dynamics. The second peak on the left of the growth rate distribution represents a minority of cells growing at slow growth rate, while the peak on the right corresponds instead to the majority of cells growing at the maximal growth rate.

Emergence of biphasic killing in the weakly non-ergodic regime

It has been demonstrated in many systems that antibiotic-mediated killing is proportional to growth-rate [15]. We then describe cell killing by the rate $k(p)$, given by $k(p(t)) = k_0 g(p(t))$, where $k_0$ is a proportionality constant that quantifies the degree of growth-rate dependency.

In the weakly non-ergodic phase the total population $X(t)$ can be regarded again as a continuum of subpopulations each labeled by the protein content,

$$\frac{dX_p}{dt} = -k_{eff}(p) X_p(t) \quad \text{with} \quad k_{eff}(p) = \frac{\mu_0(k_0 - 1)}{1 + \kappa p}, \quad p \in [0, \infty] \quad (14)$$

where $0 \leq k_0 < 1$ identifies a growth process, while $k_0 > 1$ represents antibiotic exposure killing. As before, the set of equations (14) corresponds to the picture of non-interacting populations derived by our approximation of weakly non-ergodic regime valid for large but finite observational times.

Under this approximation, the total population can be obtained by integrating over $p$,

$$X(t) = \int_0^\infty dp\, X_p(t) = \int_0^\infty dp\, X_p(0)\, e^{-k_{eff}(p)t}, \quad (15)$$

where we used the solution of (14) assuming $p$ to be independent of time, consistent with the slow fluctuations limit. By multiplying and dividing by $X(0)$, and using the definition $w(p,t) = X_p(t) / X(t)$ at time $t$, we immediately obtain

$$X(t) = X(0) \int_0^\infty dp\, w(p,0)\, e^{-k_{eff}(p)t}, \quad (16)$$

with $w(p,0)$ the protein probability distribution at the beginning of antibiotic exposure. This is the so-called static disorder approximation [32], which is indeed an approximation of the exact time dependent dynamics, but nonetheless captures well the biphasic features of the antibiotic killing in the regime of weak ergodicity breaking.

Notice that in absence of antibiotic, for $k_0 = 0$, the long-time dominant contribution in the static disorder approximation, Eq. (16), comes from the zero protein exponential, with a divergent weight, and therefore Eq. (16) reduces to describing cells growing at the maximal growth rate. The opposite situation is realized when $k_0 > 1$, since the slower exponential decays, with $p$ large, are the ones that dominate at long times.

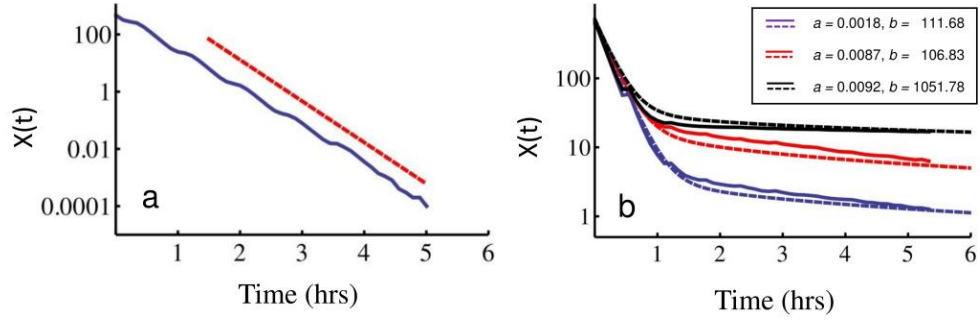

**Fig. 5. Killing curves showing the phenomenon of persistence.** Killing curves result from direct Gillespie simulations and are here compared with the static disorder approximation, Eq. (16). **(a)** Ergodic regime. Parameters are the same as in Fig. 1, with $k_0$=5. In this case, no biphasic behaviour is apparent. The dashed red line corresponds to the slope of a single exponential killing with $k_{eff} = \mu_0(k_0-1)/(1+\kappa\langle p\rangle)$ and $\langle p \rangle = ab = 43.2$ nM. Panel **(b)** shows different parameter sets characterizing the weakly non-ergodic regime. Parameter sets for the simulation were: (**Red**) $k_1$=2.0·10$^{-7}$, $k_2$=1.0, $\gamma_1$=0.01, $\gamma_2$=4·10$^{-5}$ (all units in sec$^{-1}$), $T_0$=2100 (sec), $\kappa$=1.0 (nM)$^{-1}$, $V_0$=1.7 fl, $k_0$=5; (**Blue**) $k_1$=1.0·10$^{-6}$, $k_2$=1.0, $\gamma_1$=0.01, $\gamma_2$=4·10$^{-5}$ (all units in sec$^{-1}$), $T_0$=2100 (sec), $\kappa$=1.0 (nM)$^{-1}$, $V_0$=1.7 fl, $k_0$=5; (**Black**) $k_1$=1.0·10$^{-6}$, $k_2$=10.0, $\gamma_1$=0.01, $\gamma_2$=4·10$^{-5}$ (all units in sec$^{-1}$), $T_0$=2100 (sec), $\kappa$=1.0 (nM)$^{-1}$, $V_0$=1.7 fl, $k_0$=5. The values for the mean number of bursts $a$, and for the mean burst size $b$ were fitted from the corresponding protein distribution and used to evaluate the static disorder approximation (16), indicated with dashed lines. The corresponding values of $a$ and $b$ are reported in the legend box for ease of reading. Notice the regularly spaced jolts, more apparent during the fast killing phase, corresponding to the majority of cells dividing at regular intervals $T_0$. The biphasic behaviour of the killing curve depends qualitatively on both $a$ and $b$. The lower the mean number of bursts $a$, the longer the initial killing phase, and the smaller the persister population, while the larger the mean burst size $b$, the flatter the persister tail. In general, within the present model persistence requires small $a$'s and large $b$'s.

We also performed explicit simulations of antibiotic killing. When $a \ll 1$ and $b \gg 1$, we find that the population killing curve shows a clear biphasic behavior (Fig. 5(b)): an initial exponential killing is followed by a slower tail representing killing of cells at a much lower rate. This is in contrast to the ergodic regime, where no sign of biphasic behaviour is apparent (Fig. 5(a)). In Fig. 5(b), we also show that, for fixed mean burst size $b$, decreasing the mean number of bursts $a$ produces a qualitative increase of the biphasic behavior; while for fixed $a \ll 1$, increase of the mean burst size $b$ causes the slow tail to become flatter. These simulation results fit very well with the prediction from the static disorder approximation, Eq. (16). In this case the parameters $a$ and $b$ were estimated as described above from the protein distribution, and fed into the static disorder approximation (16). No other fitting parameters were required.

## Discussion

Understanding the emergence of different phenotypes in clonal populations is a fundamental issue in cell biology that is relevant to many biomedical phenomena.

The ubiquitous existence of gene feedback, and more in general non-linear gene regulation, certainly plays a role in setting the stage [8]. For instance, in the context of understanding the role of stochasticity in cell-to-cell communications by quorum sensing, the Authors of [33] show by analysis and simulations how unimodal and bimodal distributions of signaling molecules can emerge for different values of the diffusion coefficient. This result descends remarkably only from the interplay of transcriptional noise and diffusional processes. Gene feedback circuits provide in general single-cell multistability, which is the first ingredient for realizing population heterogeneity in genetically identical cells. The epigenetic landscape, characterized by hills and valleys, is a useful pictorial representation of these dynamics.

However, the identification of different metastable states, and their basins of attraction, is not enough by itself to account for the emergence of different phenotypes. Stochastic processes allow cells to explore all possible available states, and may mask the underlying dynamics, by making the system hop quickly from state to state. What we mean and measure as a specific phenotype relies instead on the idea that the fluctuations responsible for state hopping must be slow enough for cells to maintain a biological individuality over typical observational times. In the epigenetic landscape picture, cells need to perform slow transitions among the different available valleys, so as to become in principle observable while spending time in any of them.

There may be multiple sources of static heterogeneities in the population. Rugged landscapes [11], extrinsic noise, such as heterogeneity in the number of ribosomes or RNA polymerases [4], or diffusional processes [33], are among them. In this paper we instead propose that slowdown of protein fluctuations can in fact produce stable heterogeneities in the population. In particular, key to our results is the introduction of protein controlled division times at the single cell level, which effectively acts as a mechanism that reduces the efficiency of protein mixing during cell division [14].

As a result, the phenomenon of ergodicity breaking takes place. Ergodicity breaking is a concept that is borrowed from the physical and mathematical sciences, where it plays a major role in dynamical systems theory and statistical mechanics. It has been introduced already in the Biology literature for instance in [2] to account for non-genetic variability in the evolution of cancer cells. Here we revisit the concept by introducing the related notion of weak ergodicity breaking, which we show to be responsible for the emergence of growth rate phenotypes. However, we suggest that this notion can actually be more general, and may offer a general way of linking temporal noise at the single cell level to static heterogeneities at the population level. Our definition of weak ergodicity breaking relies on the observational time being finite. However, if the system is characterized itself by an infinite relaxational time, the definition of weak ergodicity breaking can be extended to include infinite observational times [9,10]. The pictorial description of this intriguing situation is that the phase space would be connected, but it would take an infinite time for the system to explore it, and therefore to equilibrate. This specific situation may in fact be realized, either by the types of systems investigated here, or more in general in systems exhibiting inverse power law relaxational dynamics, characterizing often processes with memory.

The emergence of weakly non-ergodic components can account for the phenomenon of bacterial persistence. We here extend our growth model to include the effect of bactericidal agents, and show how the resulting dynamics is consistent with most or all of the available data on persistence. Firstly, it is entirely consistent with the established link between increased level of persistence and slow-growing and starved cells [1]. The model is also consistent both with the observation that overproduction of any gene which slows growth appears to increase persistence [34]; and the finding that a plethora of genes and mechanisms can modify persistence levels [35–38]. It is also consistent with the failure to construct/identify regulatory mutants that exist in either pure persister or non-persister states; since there is no regulatory circuit driving the transition between states.

The ergodicity-breaking model of persistence is distinctive in that it requires neither 'persistence genes', nor 'persistence states'. The model has many interesting implications for the evolution and maintenance of persistence. Both the mean number of bursts $a$ and the mean burst size $b$ are potentially evolvable parameters whose values, at the level of the individual gene, will influence the distribution of growth rate and drug tolerance, at the level of population. We note that, for the class of models analyzed here, (with growth controlled by an inhibitory gene), high rates of persistence are optimally achieved by placing growth rate under the control of an inhibitory gene that is transcribed at low levels and translated at high levels. However, our model is general, such that tuning of any gene controlling negatively growth rate will potentially be capable of modifying persistence levels.

We emphasize that our model does not exclude other mechanisms contributing to persistence. In our model the emergence of persistence is not genetically regulated. We do not assume the existence of mechanisms that by reacting to environmental conditions activate (or deactivate) synthesis of growth controlling proteins. Even though such mechanisms may be in place, we make instead the hypothesis that these are not necessary to explain persistence. Our view is that the population of persisters, pre-existing to antibiotic exposure, is anyway present because of stochastic fluctuations of any growth-inhibiting protein, and is not related to the specific regulated tuning of the expression of any specific gene. In this respect, subpopulations of normal and persister cells emerge naturally as a consequence of the growth phenotypic heterogeneity resulting from the mechanism of ergodicity breaking. We also believe that our model has significant implications for efforts to develop novel strategies to more efficiently kill, or prevent the formation of, persister cells in infectious disease and the environment.

Finally, the model of ergodicity breaking as an engine for driving growth rate heterogeneity may be more general, and have wider implications for our understanding of the emergence of cellular phenotypes. Cell growth rate is an important parameter determining response of cells to a range of stresses, signaling molecules and drugs, such as cancer chemotherapeutic agents. Indeed, cancer chemotherapy demonstrates a very similar phenomenon to bacterial persistence: a subpopulation of genetically identical but drug-tolerant cells [39]; which may thereby be driven by a similar mechanism as the model of bacterial persistence described here. Moreover, emergence of weakly non-ergodic components is not necessarily restricted to growth rate control but may be a more general mechanism for the emergence of distinct cellular phenotypes in isogenic populations.

## Methods
Analysis of the growth rate distribution

Here we show that the growth rate distribution (13) presents two modes for $a < 1$ and $(1+a+1/\kappa b)^2 - 8/\kappa b > 0$.

At high growth rates, for $\mu$ close to $\mu_0$, the behavior of the growth rate distribution (13) depends on the mean number of bursts $a$. For $a > 1$, it is straightforward to verify that $\chi(\mu) \to 0$ for $\mu \to \mu_0$. For $a < 1$, we have instead:

$$\chi(\mu) \approx \frac{1}{\mu_0 (\kappa b)^a \Gamma(a)} \left(\frac{\mu_0}{\mu} - 1\right)^{a-1} \to \infty \quad \text{for} \quad \mu \to \mu_0 \text{ and } a < 1, \qquad (17)$$

which shows that a peak appears at $\mu = \mu_0$. This peak is expected, because it corresponds to the peak at $p \approx 0$ of the Gamma distribution for $a < 1$, accounting for the majority of cells presenting negligible protein concentration, and growing therefore at the maximal growth rate.

The behavior at low growth rates of the growth rate distribution (13) is also simple to compute. In order to search for non-monotonic behavior of the function $\chi(\mu)$, and for other modes, we can carry out the first derivative of (13), and compute its roots. The calculation is straightforward, and leads to the two following values:

$$\mu = \frac{\mu_0}{4} \left[ \left(1 + a + \frac{1}{\kappa b}\right) \pm \sqrt{\left(1 + a + \frac{1}{\kappa b}\right)^2 - \frac{8}{\kappa b}} \right]. \qquad (18)$$

For these to be distinct and real, we need to require the discriminant to be positive, namely:

$$\Delta(a,b) = \left(1 + a + \frac{1}{\kappa b}\right)^2 - \frac{8}{\kappa b} > 0. \qquad (19)$$

When this is the case, the roots will also be both positive. Since the function $\chi(\mu)$ is non-negative in $[0, \mu_0]$, with $\chi(0) = 0$, this shows that for any combination of values $a$ and $b$ with $a < 1$ and satisfying the reality condition (19), the function $\chi(\mu)$ presents a maximum, followed by a minimum, followed by the divergent value identified by (17). This implies that further to the mode at $\mu = \mu_0$, another mode is present at low growth rates, and is identified by (18) when $a < 1$ and (19) is verified. We show in Fig. 6 a contour plot of Eq. (19) for $\kappa = 0.01$, which identifies the combined values of the mean number of burst $a$ and the mean burst size $b$ such that two modes are present. It is interesting to note that two regions of $b$ allow for two modes to be present, namely either $b$ fairly small or $b$ fairly large. However the region characterized by very small $b$ shows in general tiny slow growth peaks (data not shown), while the region with large $b$ presents more pronounced peaks. This is consistent with the picture that the parameter $b$ "separates" the ergodic components, pushing cells either in the slow growing state, or in the fast growing one.

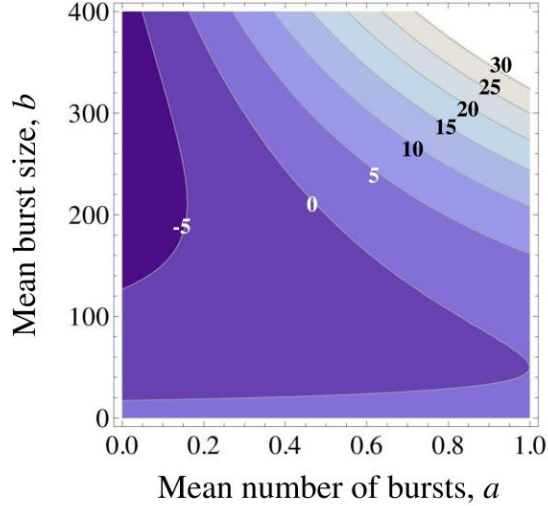

**Fig. 6. Contour plot showing the discriminant $\Delta(a,b)$ given by Eq. (19).** In this plot $\kappa = 0.01$ (nM)$^{-1}$ has been assumed. Weakly non-ergodic behaviour, characterized by the emergence of two modes in the growth rate distribution, is predicted for $a$ and $b$ values such that $\Delta(a,b)$ is positive.

### Exact Stochastic Simulations

The simulation of the model proposed here is based on an extension of the Gillespie algorithm [30] so as to incorporate protein dependent growth rates. The two-stage gene expression model detailed in the text, which includes transcription, translation and protein and mRNA degradation, has been simulated in standard fashion following [30]. Instead protein controlled cell division is non-standard, and has required a specific modification of the algorithm.

We assume that at cell division all molecular species, except DNA elements, are binomially split between two daughter cells. In order to compute the instant of cell division, we monitored the cell volume by using the following expression:

$$V_2 = \frac{\kappa n}{W\left[\kappa n \exp\left(\frac{\kappa n}{V_1} - \log V_1 - g_0 \tau\right)\right]}. \qquad (20)$$

In this expression $W$ is the Lambert function (see next section), $n$ is the number of protein molecules, $\tau$ is the Gillespie time, and $V_1$ and $V_2$ are respectively the cellular volume at the previous and at the present Gillespie iteration. We impose division if $V_2 > 2V_0$, where $V_0$ is the cellular volume just after cell division. In our simulations we used $V_0 = 1.7 \cdot 10^{-15}$ litres. The Lambert function involved in Eq. (20) was computed numerically by using the f77 subroutine `am05_xscss_lambertw` [40], downloadable from [http://dft.sandia.gov/functionals/AM05.html].

If a division occurs, we reset the simulation time $t$ at the value $t + \Delta t$, where

$$\Delta t = \frac{\kappa n}{g_0}\left(\frac{1}{V_1} - \frac{1}{2V_0}\right) + \frac{1}{g_0}\log\left(\frac{2V_0}{V_1}\right). \tag{21}$$

This expression is also derived in the next section.

The rest of the algorithm, involving transcription, translation and degradation processes is standard as from [30].

### The Lambert Function

The law of volume growth with protein control used in our simulations can be derived in terms of Lambert functions. The Lambert function is defined in general as the solution of the equation [41]:

$$z = W(z)\exp(W(z)). \tag{22}$$

In what follows, we will limit ourselves to the real-valued Lambert function.

Let us consider Eq. (2) and make it explicit in the volume variable:

$$\frac{dV(t)}{dt} = \frac{g_0}{1 + \kappa\dfrac{n(t)}{V(t)}} V(t). \tag{23}$$

In order to solve this equation we need knowledge about the time dependency of protein molecules numbers $n(t)$. However, during the Gillespie time $\tau$, or more in general for times over which the protein content does not change, we can replace $n(t) = n$ in (23), and solve the equation formally. By setting $\tilde{V}_i = 1/V(t_i)$ for $i = 1,2$, it is straightforward to rewrite (23) as

$$\tilde{V}_2 \exp(\kappa n \tilde{V}_2) = \tilde{V}_1 \exp(\kappa n \tilde{V}_1 - g_0(t_2 - t_1)), \tag{24}$$

whose formal solution can be expressed in terms of the Lambert function as

$$\tilde{V}_2 = \frac{1}{\kappa n} W\left[\kappa n \tilde{V}_1 \exp(\kappa n \tilde{V}_1 - g_0(t_2 - t_1))\right]. \tag{25}$$

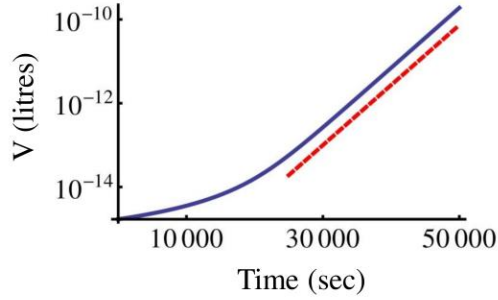

**Fig. 7. Behaviour of the volume growth law.** Plot of Eq. (26) for $t_1 = 0$ and $V(t_1) = V(0) = V_0$ (blue curve). Parameters were $V_0 = 1.7 \cdot 10^{-15}$ litres, $\kappa = 0.01$ (nM)$^{-1}$, $g_0 = \ln 2/2100$ (sec$^{-1}$), and the protein copy number was rescaled to $n = 500 \cdot 10^9 / N_A$ with $N_A = 6.022 \cdot 10^{23}$ (Avogadro number). The dashed red line represents the slope associated with the exponential asymptotic growth law $\exp(g_0 t)$, shown for comparison. The shift between the two curves is due to the arbitrary prefactor in front of the exponential. The Lambert function introduces a deviation from exponential volume growth at short times.

From this we immediately obtain

$$V(t_2) = \frac{\kappa n}{W\left[\kappa n \exp\left(\frac{\kappa n}{V(t_1)} - \log V(t_1) - g_0(t_2 - t_1)\right)\right]}, \quad (26)$$

which leads to Eq. (20). In Fig. 7 we show a numerical evaluation of the volume growth law (26) for $n = 500$. It is interesting to note that the asymptotic behaviour of the volume growth law (26) is exponential, with a rate given by $g_0$. In Fig. 7 we also plot this asymptotic behaviour for comparison. This shows how the growth law assumed here, eq. (26), deviates from the standard exponential behaviour only at short times.

Eq. (21) can be easily obtained by imposing

$$2V_0 = \frac{\kappa n}{W\left[\kappa n \exp\left(\frac{\kappa n}{V} - \log V - g_0 \Delta t\right)\right]}, \quad (27)$$

and using the definition of Lambert function (22). In eq. (27), $V$ is the last recorded value of the cellular volume.

## Acknowledgements
We acknowledge support from The Wellcome Trust (ref: 088677).